\documentclass[structabstract]{aa}  
\usepackage{graphicx}
\usepackage{txfonts}

\usepackage{natbib}
\bibpunct{(}{)}{;}{a}{}{,} 

\newcommand{\ind}[1]{_{\mathrm{#1}}}
\newcommand{\ex}[1]{^{\rm #1}}

\newcommand\deltanunu{\Delta\nu(\nu)}
\newcommand\HBR{\mathcal{R}}
 \def\ptt{\hbox{$10^{-4} I_{\rm c}$}}

\begin{document}
   \title{HD~46375: seismic and spectropolarimetric analysis of a young Sun hosting a Saturn-like planet\thanks{The CoRoT space mission, launched on 2006 December 27, was developed and is operated by the CNES, with participation of the Science Programs of ESA, ESA's RSSD, Austria, Belgium, Brazil, Germany and Spain.}}

   \author{P.~Gaulme\inst{1}
          \and S.~Deheuvels\inst{2}
          \and W.~W.~Weiss\inst{3}
          \and B.~Mosser\inst{2}
          \and C.~Moutou\inst{4}
          \and H.~Bruntt\inst{2}
           \and J.-F.~Donati\inst{5}
          \and M.~Vannier\inst{6}
          \and T.~Guillot\inst{7}
          \and T.~Appourchaux\inst{1}
          \and E.~Michel\inst{2}
         \and M.~Auvergne\inst{2} 
         \and  R.~Samadi\inst{2}
                  \and F.~Baudin\inst{1} 
         \and C.~Catala\inst{2}
          \and A.~Baglin\inst{2}
          }

    \institute{$^1$ Institut d'Astrophysique Spatiale, UMR 8617, Universit\'e Paris Sud, 91405 Orsay Cedex, France \\
   $^2$ LESIA, UMR 8109, Observatoire de Paris, 92195 Meudon Cedex, France\\
$^3$ University of Vienna, Inst of Astronomy, T\"urkenschanzstr. 17, AT 1180 Vienna, Austria\\
$^4$ Laboratoire d'Astrophysique de Marseille, 38 rue Fr\'ed\'eric Joliot-Curie, 13388 Marseille Cedex 13, France\\
$^5$ LATT, UMR 5572, CNRS and University P. Sabatier, 14 Av. E. Belin, 31400 Toulouse, France\\
$^6$  Laboratoire Fizeau, Universit\'e de Nice, CNRS-Observatoire de la C\^ote d'Azur, 06108 Nice Cedex 2, France\\ 
$^7$  Laboratoire Cassiop\'ee, Universit\'e de Nice, CNRS-Observatoire de la C\^ote d'Azur, 06304 Nice Cedex 4, France\\ 
              \email{Patrick.Gaulme@ias.u-psud.fr}
        }
        

 
  \abstract
{HD~46375 is known to host a Saturn-like exoplanet orbiting at 0.04 AU from its host star. Stellar light reflected by the planet was tentatively identified in the 34-day CoRoT run acquired in October-November 2008. }
{We constrain the properties of the magnetic field of HD~46375 based on spectropolarimetric observations with the NARVAL spectrograph at the Pic du Midi observatory. In addition, we use a high-resolution NARVAL flux spectrum to contrain the atmospheric parameters. With these constraints, we perform an asteroseismic analysis and modelling of HD~46375 using the frequencies extracted from the CoRoT light curve.}
{ We used Zeeman Doppler imaging to reconstruct the magnetic map of the stellar surface. In the spectroscopic analysis we fitted isolated lines using 1D LTE atmosphere models. This analysis was used to constrain the effective temperature, surface gravity, and chemical composition of the star. To extract information about the p-mode oscillations, we used a technique based on the envelope autocorrelation function (EACF).}
{From the Zeeman Doppler imaging observations, we observe a magnetic field of $\approx5$~gauss. From the spectral analysis, HD~46375 is inferred to be an unevolved K0 type star with high metallicity ${\rm [Fe/H]}=+0.39$. Owing to the relative faintness of the star ($m\ind{hip}=8.05$), the signal-to-noise ratio is too low to identify individual modes. However, we measure the p-mode excess power and large separation $\Delta \nu_0=153.0\pm0.7\,\mu$Hz.}
{{We are able do constrain the fundamental parameters of the star thanks to spectrometric and seismic analyses. We conclude that HD~46375 is similar to a young version of $\alpha$~Cen~B. This work is of special interest because of its combination of exoplanetary science and asteroseismology, which are the subjects of the current {\it Kepler} mission and the proposed {\it Plato} mission.}}
   {}

\maketitle

%

\section{Introduction}
The CNES CoRoT satellite has been orbiting the Earth at an altitude of 896 km, in a polar orbit, since late December 2006. Its scientific goals are divided into two themes: asteroseismology and the detection of transiting exoplanets. Apart from the 5-month long observation runs dedicated to the main targets, CoRoT performs short runs intended to observe a large set of stars with different properties. We present the observation of the non-transiting extrasolar system HD 46375 in the astero-seismological field of CoRoT, during a 34-day run between October 9 and November 11 2009. The objective is to obtain observational constraints on both the star and the planet, by performing a seismic analysis of the star and detecting the changing phases of the planet during its orbital period. We already know that this planet does not transit in front of the star, which implies that the inclination of the planetary orbital plane is less than 83$^\circ$ \citep{Henry_2000}. However, we have discovered that the light curve, which is the sum of the star emission and the light reflected by the planet, is modulated by the reflected planetary flux (Gaulme et al 2010b, Paper II). This paper is dedicated to the analysis of the stellar properties, by means of the seismic analysis of the CoRoT data and spectro-polarimetric ground-based support. An analysis of the CoRoT light curve that is focused on the planet will be presented in Paper II.

HD~46375 is a solar-like star with properties similar to $\alpha$~Cen~B that is particularly interesting because it may be a younger version of the $\alpha$~Cen~A$+$B system. It is an unevolved K0~type star of high metallicity, with apparent magnitude from Hipparcos database $m\ind{hip} = 8.05$. Moreover, from the rotation-activity relationship \citep{Noyes_1984} and given the measurements of Ca\,{\sc ii} H and K emission $\log R\ind{H,K}' = -4.94$ and $(B-V) = 0.86$ reported in \citet{Marcy_2000}, we estimate the stellar rotation period at $42^{+9}_{-7}$ days. The star is understood to be very quiet with relative variations of the order of a few 100 ppm on time scales of a few days. 

Since early 2007, CoRoT has been observing solar-like oscillations in several solar-like stars, subgiants, and giants. However, the number of late-type main-sequence targets is still small. Most of the main targets have been F type stars \citep{Barban_2009, Garcia_2009, Mosser_2009_175726, Benomar_2009_HD49933}. Photometric observations with CoRoT favour F-type stars, since they are typically brighter and also expected to exhibit higher amplitude oscillations. Asteroseismic spectroscopic ground-based campaigns are limited to bright stars ($V<6$), and some G and early K type stars have been observed, e.g. $\iota$ Hor \citep{Vauclair_2008}, $\alpha$~Cen~B \citep{Kjeldsen_2005}, $\tau$~Ceti \citep{Teixeira_2009}, and HD~49385 (Deheuvels et al. 2010, A\&A, submitted). With HD~46375, we have the opportunity to investigate solar-like oscillations in a low-mass K0-type star, to improve the determination of its internal structure and to more tightly constrain the fundamental parameters of both the stellar and planetary system.

To complement the CoRoT data, ground-based observations were carried out at the Pic du Midi observatory. We used the NARVAL spectropolarimeter to obtain both a high resolution visible spectrum and magnetic maps of the stellar surface, by means of Zeeman Doppler imaging. These observations allow us to monitor the changes in magnetic activity.

We first present the analysis of the spectropolarimetric and spectrometric data (Sect.~2). We then present the CoRoT lightcurve (Sect.~3) and the seismic analysis (Sect.~4). We then combine the constraints to develop an asteroseismic model of HD~46375 (Sect.~5). Finally, we discuss the implications for the properties of the star-planet system (Sect.~6). 

\section{Magnetic imaging and stellar parameters}
\subsection{Spectropolarimetric observations}

\begin{figure}[t]
\includegraphics[width=9cm]{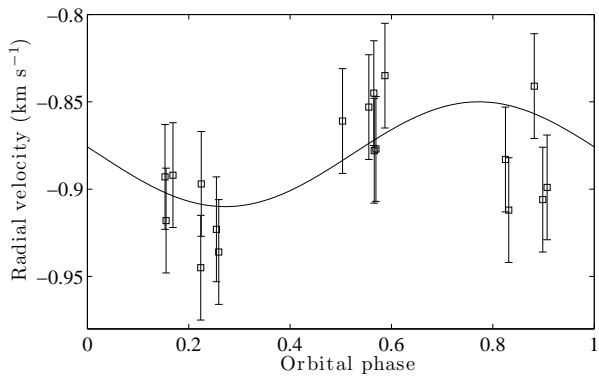}
\caption[]{Radial velocity associated with each exposure of the Sept-Oct 2008 NARVAL observations of HD 46375. The measurements are folded over the planetary orbital phase (3.02357 days), estimated by \citet{Butler_2006}. }
\label{fig:rv}
\end{figure}

\begin{figure}[t]
\centering
\rotatebox{-90}{\includegraphics[height=7cm]{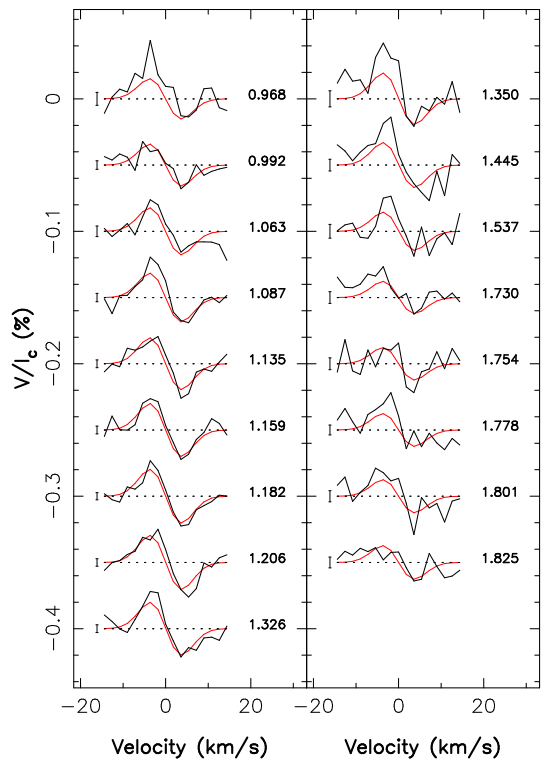}}
\caption{Maximum-entropy fits (thin red line) to the Stokes V LSD profiles observed by TBL/NARVAL (thick black line) of HD 46375. The rotational phase (in radians) of each observation and 1$\sigma$ error bars are also shown next to each profile.}
\label{fig:ZD_fit}
\end{figure}

\begin{table}
  \caption{Journal of observations. From left to right: heliocentric Julian date at mid-exposure (Col. 1);
total exposure time (Col. 2); peak signal-to-noise ratio (per 2.6~km s$^{-1}$
velocity bin) of each four-subexposure polarisation sequence (Col. 3); rms noise level (relative to the unpolarized continuum 
level $I_{\rm c}$ and per 1.8~km s$^{-1}$ velocity bin) in the circular 
polarization profile produced by least squares deconvolution (LSD) (Col. 4); orbital cycle (Col. 5); rotational cycle (Col. 6); and the radial velocity associated with each observation (Col. 7). 
  \label{tab_param}}
  \scalebox{0.85}{%
\begin{tabular}{c  c  c  c  c  c  c  c c}
\hline
 HJD          & $t\ind{exp}$ &  SNR  & $\sigma\ind{LSD}$ & $\phi\ind{orb}$ & $\phi\ind{rot}$ & RV\ \\
\tiny{(2,454,000+)} & \tiny{(s)} &        &   \tiny{(\ptt)}           & \tiny{(1,200+)}  &  \tiny{(5+)}   & \tiny{(km s$^{-1}$)} \\
\hline 
730.67708 & $4\times800$ & 320 & 1.0 & 10.263 & 0.969 & -0.936 \\
731.66854 & $4\times900$ & 480 & 0.6 & 10.591 & 0.992 & -0.835 \\
734.63876 & $4\times900$ & 400 & 0.8 & 11.573 & 1.063 & -0.877 \\
735.65936 & $4\times900$ & 500 & 0.6 & 11.911 & 1.087 & -0.899 \\
737.65226 & $4\times900$ & 530 & 0.6 & 12.570 & 1.135 & -0.878 \\
738.65755 & $4\times900$ & 510 & 0.6 & 12.902 & 1.159 & -0.906 \\
739.64068 & $4\times900$ & 550 & 0.5 & 13.227 & 1.182 & -0.945 \\
740.64321 & $4\times900$ & 530 & 0.6 & 13.559 & 1.206 & -0.853 \\
745.69059 & $4\times900$ & 510 & 0.6 & 15.228 & 1.326 & -0.897 \\
746.72039 & $4\times400$ & 250 & 1.2 & 15.569 & 1.350 & -0.845 \\
750.70071 & $4\times900$ & 350 & 0.9 & 16.885 & 1.445 & -0.841 \\
754.55168 & $4\times650$ & 290 & 1.0 & 18.159 & 1.537 & -0.918 \\
762.64276 & $4\times650$ & 360 & 0.9 & 20.835 & 1.730 & -0.912 \\
763.66335 & $4\times650$ & 360 & 0.9 & 21.173 & 1.754 & -0.892 \\
764.67493 & $4\times650$ & 420 & 0.7 & 21.507 & 1.778 & -0.861 \\
765.64663 & $4\times650$ & 390 & 0.8 & 21.828 & 1.801 & -0.883 \\
766.63943 & $4\times650$ & 390 & 0.8 & 22.157 & 1.825 & -0.893 \\
\hline 
\end{tabular}
}
\end{table}

\begin{figure}[t]
\includegraphics[width=5cm]{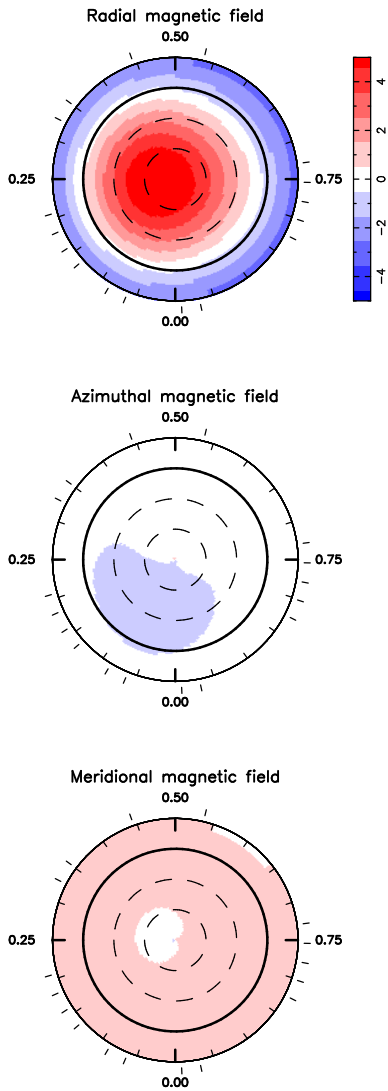}
\caption{Maximum-entropy reconstruction of the large-scale magnetic topology of HD 46375 as derived by the TBL/NARVAL data set. The radial, azimutal, and meridional components of the field are shown from top to bottom. Magnetic flux values are given in gauss as indicated along the vertical colour bar. The star is shown in flattened polar projection down to latitudes of $-30^\circ$, with the equator depicted as a bold circle and parallels as dashed circles. Radial ticks around each plot indicate the rotational phases of observations.}
\label{fig:ZD_map}
\end{figure}

The spectropolarimetric data were acquired with the NARVAL spectropolarimeter, permanently mounted at the 2-m Bernard Lyot Telescope at the Pic du Midi Observatory (France). Eighteen spectra were collected during a 35-night run between September 21 and October 27, 2008. A single spectrum covers the full optical domain from 3700 to 10\,480~\AA\ at a spectral resolution of 65\,000. The circular polarization is measured in a four-exposure sequence with different configurations of the waveplates. The data were processed with the Libre-Esprit software \citep{Donati_1997, Donati_2007}.  

We calculated the stellar rotational and planetary orbital phase corresponding to our observations, by assuming a stellar rotation period of 42 days, as obtained from the rotation-activity relationship, and the planetary orbital period from \citet{Butler_2006}\footnote{\vspace{-0.3cm}\begin{eqnarray}
HJD &=& 2454480.0   + 42.0\  \phi\ind{rot}\\
HJD &=& 2451071.359 + 3.023573\  \phi\ind{orb}
\end{eqnarray} }.
Least squares deconvolution (LSD) techniques were applied to the spectra to improve the signal-to-noise ratio (SNR), with a  K0 template mask \citep{Donati_1997}. More than 6000 stellar lines were used to derive mean profiles in the Stokes I and V parameters. The mean I profile was used to derive the radial velocity (Fig. \ref{fig:rv}), whose variation is compatible with \citet{Butler_2006}. The error in the radial velocity (about 30 m s$^{-1}$ per individual measurement) and the poor orbital sampling (with a period close to an integer number of days) do not allow us to refine the planetary ephemeris, since the semi-amplitude of the radial-velocity variation is only 34 m s$^{-1}$. We present the journal of observations in Table \ref{tab_param}.

\subsection{Magnetic imaging}
We use Zeeman Doppler imaging (ZDI) to reconstruct the magnetic map of the stellar surface. ZDI is a tomographic imaging technique based on the inversion of the time series of the Stokes V profiles into a map of the magnetic topology \citep{Donati_Brown_1997,Donati_2006}. An example of this modelling is given by \cite{Fares_2009}, who applied it to the planet-bearing star $\tau$ Boo. 

The LSD Stokes V signature is detected in two thirds of the spectra, with small variations in the polarized profile (Fig. \ref{fig:ZD_fit}). Those variations over 35 nights are limited; in particular, the field component along the line of sight is always  positive and does not switch sign across the rotation cycle, suggesting that the large-scale field topology is fairly simple and mostly  axisymmetric with respect to the rotation axis of the star. We note that we checked the V Stokes profiles are significantly detected. Each profile shown in Fig.~\ref{fig:ZD_fit} has indeed a flat counterpart in the null profile, used for check, and no flat noise pattern was observed in the present data. 

The fact that the magnetic Stokes V measurements do not exhibit any periodicity, confirms that the rotation period is longer than 35 days.
Therefore, the temporal coverage of the ZDI observations is shorter than the expected rotation period. 
Hence, it is not possible to constrain either the rotation period or the differential rotation at the stellar surface. We use the $42^{+9}_{-7}$ day estimate of the rotation period to invert the data. In addition, from the stellar modelling performed thanks to the seismic and spectrometric analysis (Sect. \ref{modele}), the stellar radius is $0.91\pm 0.02$~R$\ind{\sun}$. Hence, the rotational velocity at the equator is $v =1.1 \pm 0.2$ km s$^{-1}$. In Sect. \ref{HR_spectro}, the projected rotation velocity is estimated to be $v \sin i = 1.2 \pm 0.4$ km s$^{-1}$. This estimate is not fully compatible with the equatorial velocity estimate, whose upper limit is 1.3 km  s$^{-1}$ this also being the $v \sin i$ upper limit. Hence, $v \sin i$ lies in the range $[0.8,1.3]$ km s$^{-1}$. We then have a rough estimate of the stellar inclination: by assuming Gaussian distributions for $v\sin i $ and $v$, we obtain an estimate of the inclination angle $i = 50\pm18^\circ$.      

For simplicity, we modelled the Stokes V profiles by taking an inclination of $i=45^\circ$ (well within the error bars) and a line-of-sight projected equatorial rotation velocity $v \sin i = 0.9$ km s$^{-1}$. The magnetic topology derived with ZDI is shown in Fig. \ref{fig:ZD_map}, with the  corresponding maximum-entropy fit to the data being shown in Fig. \ref{fig:ZD_fit}. The reconstructed field structure resembles a simple dipole slightly tilted, by about $15^\circ$, with respect to the rotation axis. The magnetic strengh at the pole is about 5 gauss. 
We note that the result is only moderately sensitive to the true value of $i$ and changes by about $\pm10^\circ$ produce very small changes in the corresponding magnetic image. The stellar inclination is an input of the magnetic modelling and not a fitted output parameter. In other words, the magnetic mapping does not allow us to constrain the stellar inclination. 
 
The magnetic structure of HD~46375 appears to be significantly different from those of other planet-bearing stars such as $\tau$ Boo \citep{Catala_2007, Donati_2008, Fares_2009} and HD~189733 (\citealt{Moutou_2007}, Fares et al., submitted to MNRAS). HD~46375 presents a far more simple magnetic structure. More specifically, the magnetic field of these 2 stars shows a significant or even mainly toroidal component whereas that of HD 46375 is predominantly poloidal. This difference mostly reflects the differences in rotation periods and more precisely the difference in Rossby numbers, which is related to the dynamo properties for stars of different spectral types (e.g. \citealt{Noyes_1984}). With a rotation period of 42 days, HD~46375 features a Rossby number of about 2 and naturally falls among stars featuring very simple, mostly-poloidal and axisymmetric large-scale  magnetic topologies \citep{Donati_Landstreet_2009}.

\subsection{Stellar parameters}
\label{HR_spectro}

\begin{figure}[t]
\rotatebox{90}{\includegraphics[height=9cm]{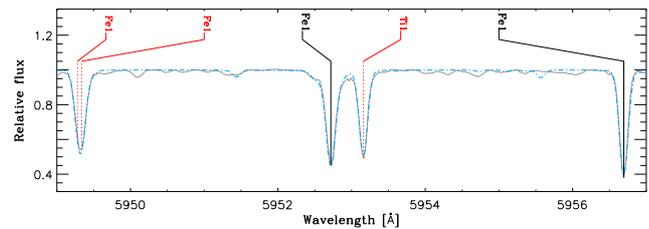}}
\caption[]{Comparison of the observed spectrum (grey line) with a synthetic profile (blue dashed-dotted line). }
\label{fig:spectrum}
\end{figure}

The high-resolution NARVAL spectra of HD~46375 were also used to determine the stellar parameters. All intensity spectra were summed prior to the spectroscopic analysis: the resulting SNR is 330 around 6000\,\AA. The summed spectrum was analysed in the spectral range $5000$ to $8500$\,\AA\ with the VWA software, which makes use of iterative fitting of synthetic profiles \citep{Bruntt_2004, Bruntt_2008}. In Fig.~\ref{fig:spectrum}, a portion of the observed spectrum is compared with its corresponding synthetic profile (dash-dotted line). We note that the star is late type (K0) and that several weak lines are not available in the adopted atomic line data from VALD \citep{Kupka_1999}. This introduces a weak bias in the abundance estimates. It may also affect the estimate of the atmospheric parameters, since weaker lines are more contaminated by unidentified lines than relatively stronger lines.

We used 159 isolated Fe\,{\sc i} lines to constrain the atmospheric parameters, by minimizing the correlations between abundance with equivalent width and excitation potential. The neutral and ionized Fe lines were required to infer the same abundance, to constrain the surface gravity to be $\log g = 4.73 \pm 0.10$ dex. Furthermore, to confirm the $\log g$ value, the synthetic profiles were compared to the Mg\,{\sc i}b lines and the broad Ca lines at 6122, 6262, and 6439\,\AA\ (\citealt{Bruntt_2010}). On the one hand, the Mg\,{\sc i}\,b lines could not be used because this region is strongly blended and affected by molecular lines for such a late-type star. On the other hand, the determined surface gravity with the three Ca lines, is, respectively, $\log g = 4.64\pm0.07$ dex, $4.63\pm0.04$ dex, and $4.69\pm0.16$ dex. These results agree with the Fe ionization balance. We computed the weighted mean value of the four estimates (ionization balance and Ca lines) as the final estimate of the surface gravity $\log g = 4.66 \pm 0.05$ dex.  

\begin{figure}[t]
\rotatebox{90}{\includegraphics[height=9cm]{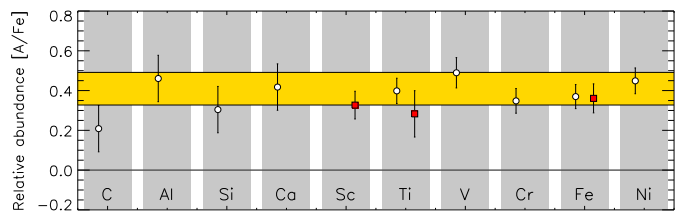}}
\caption[]{Relative abundances of 12 elements measured in HD~46375. Open circles and solid squares are mean values for neutral and singly ionized lines, respectively. The yellow horizontal bar marks the mean metallicity with 1-$\sigma$ uncertainty range.}
\label{fig:abund}
\end{figure}

The metallicity estimate is the mean value of the abundances of metallic species with at least 10 spectral lines (Si, Ti, Cr, Fe, and Ni); the relative values with respect to the Sun are shown in Fig.~\ref{fig:abund}.  We note that errors in the element abundance take into account the uncertainty in both temperature and gravity, as well as the internal scatter of abundances determined from lines of the same element. 

The most robust estimates of the effective temperature, surface gravity, and metallicity are $T\ind{eff} = 5300 \pm 60$ K, $\log g = 4.66 \pm 0.09$ dex, and ${\rm [M/H]} = +0.39 \pm 0.06$ dex. For these three parameters, systematic uncertainties of 50~K, 0.08 dex, and 0.05 dex were added. In particular, 40~K was subtracted from $T\ind{eff}$, to be consistent with the temperature scale of \cite{Bruntt_2010}. Finally, the projected rotational velocity was measured by fitting synthetic profiles to isolated lines to be $v \sin i = 2.0 \pm 0.8$\,km\,s$^{-1}$. This value is higher than the velocity estimated from \citet{Valenti_2005},  $v\sin i = 0.9\pm0.5$ km s$^{-1}$, but it still fits inside the error bars. Since both values were extracted from spectra of similar resolution (70~000 for Valenti \& Fischer and 65~000 for NARVAL), we consider the weighted mean value $v\sin i = 1.2\pm0.4$ km s$^{-1}$.

The comparison of the HD~46375 parameters with previous works shows reasonable good agreement. From spectral analysis, \cite{laws03} determined $T_{\rm eff} = 5241 \pm 44$\,K, $\log g=4.41\pm0.09$ dex, and ${\rm [Fe/H]} = +0.30 \pm 0.03$ dex. Moreover, \cite{santos04} found $T_{\rm eff} = 5268 \pm 65$\,K, $\log g=4.41\pm0.16$ dex, and ${\rm [Fe/H]} = +0.20 \pm 0.06$ dex. There is good agreement for the effective temperature, while our surface gravity and metallicity are slightly higher than the previous estimates.

\subsection{Stellar luminosity}

From the Hipparcos database, HD~46375 has an apparent magnitude  $m\ind{hip} = 8.0512\pm0.0017$. \citet{Van_Leeuwen_2007} reviewed the Hipparcos parallaxes and found that $\pi = 28.72\pm0.89$ mas, which corresponds to an absolute magnitude of  $M\ind{hip} = 5.34 \pm 0.07$ assuming that interstellar reddening is negligible.

We used bolometric corrections, calculated for the Hipparcos spectral band \citep{Cayrel_1997}, with the new values of $T\ind{eff}$, $\log g$, and [Fe/H]. The bolometric correction is equal to BC~$=  -0.30 \pm 0.03$. Therefore, the bolometric magnitude is $M\ind{bol} = 5.04 \pm 0.10$, which corresponds to a luminosity  $L = (0.68 \pm 0.06) L\ind{\sun}$. 

\section{CoRoT observations}
\subsection{Cleaning of the time series}
\label{cleaning}
As for the previous CoRoT observation runs, the time series at level ``N2'' are still contaminated by the South Atlantic Anomaly (SAA) periodic interruptions (e.g. \citealt{Auvergne_2009}). 
To diminish the SAA effect, we removed the suspect data and re-interpolated all of the gaps by replacing them with an iterative estimate of the low frequency trend of the time series, applied by \citet{Gaulme2008_a} to Jovian seismic data.  

Since these methods are not able to completely cancel the signature of the satellite orbit in the power density spectrum of the time series, we eliminate iteratively its harmonics as well as the sub-harmonics associated with the day frequency (11.57 $\mu$Hz) as done in \citet{Mosser_2009_175726}. We cleaned the power spectrum across the whole frequency range for every multiple of the satellite frequency (161.72 $\mu$Hz) and cleaned 5 daily harmonics around each harmonic.

\begin{figure}[t]
\includegraphics[width=9cm]{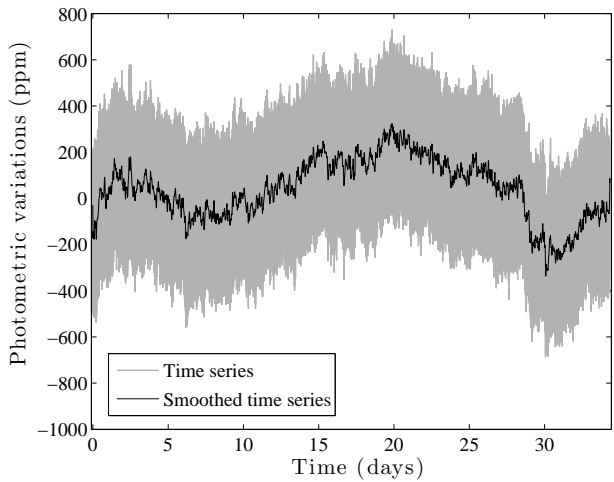}
\caption[]{CoRoT time series of HD~46375 after the re-processing described in Sect. \ref{cleaning}. The time series is plotted in grey at full resolution, and in black after a smoothing with box-car of 100-min long. The standard deviation around the low frequency trend is 176 ppm. The epoch zero corresponds to 3203.62037 days in the CoRoT calendar, that is to say after January 1, 2000 at 12:00 GMT.}
\label{vraie_lightcurve}
\end{figure}
\begin{figure}[h]
\includegraphics[width=9cm]{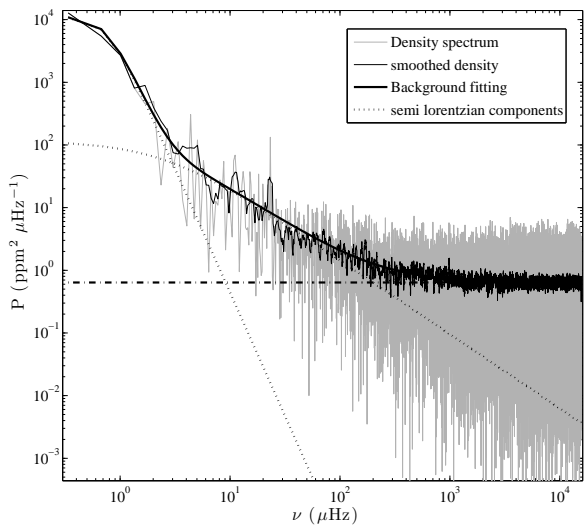}
\caption[]{Density spectrum, in log-scale axis. The black curve corresponds to the averaged spectrum obtained with a box-car with a moving weighted average, of variable width. There is no clear excess power at high frequency that could be directly related to pressure modes.
}
\label{log_log}
\end{figure}

\subsection{Analysis of the stellar activity}

The CoRoT light curve (Fig. \ref{vraie_lightcurve}) does not exhibit a high activity signature, which is also visible in the log-log representation of the power density spectrum (Fig. \ref{log_log}). It differs from previously observed F-type main-sequence stars \citep{Barban_2009,  Garcia_2009}. Compared to them, there is no clear signature at very low frequencies of the rotation period. A spot analysis performed in \citet{Mosser_2009_taches} is impossible for HD 46375. The observation time span is not long enough compared to the rotation period, and the activity level is very low, as expected for this K0V star. It is therefore impossible to identify an unambiguous signature of transiting spots. Even after a rebinning of the light curve at 1 point each CoRoT orbit, the SNR remains too low to derive a reliable rotation period. 

Furthermore, the fitting of the power spectrum with a sum of 3 semi-Lorentzian profiles, as suggested by Harvey (1985), does not converge reliably. Two components are enough to provide an acceptable fit. We fitted the power density spectrum with a maximum likelihood estimator (MLE), by taking into account the nature of the statistics of the power spectrum, which is $\chi^2$ with 2 degrees of freedom. The result is presented in Table 1. The amplitude of the power's main component is estimated to be $1.12^{+1.67}_{-0.67} \times 10^{4}$ ppm$^2\mu$Hz$^{-1}$. We note that the upper error bar is larger than the value of the amplitude; a longer run and consequently a higher frequency resolution would allow us to place tighter constraints on this parameter. The only reliable estimates are the slope $p_i$ of each semi-Lorentzian in the log-log representation.

\begin{table}[h!]
 \caption{Fitted background parameters, where $A_i$, $\theta_i$, $p_i$, and B are defined by the function $f(\nu) = B + \sum_i A_i/[1 + (\theta_i \nu)^{p_i}]$.}
\begin{tabular}{ c  c c  }
\hline
Parameter& Value & Error\\
 \hline
 A$_i$ &$1.12\ 10^4$           &   ($+$1.67/$-$0.67) $10^4$    \\        
 (ppm$^2 \mu$Hz$^{-1}$)  & $1.16\ 10^2$          &     ($+$1.97/$-$0.73) $10^2$    \\
 \hline
$\theta_i$ &   $1.31\ 10^6 $     &   ($+$0.51/$-$0.36) $10^6$       \\
(s)              &      $4.02\ 10^5 $  &   ($+$6.35/$-$2.46) $10^5$\\  
 \hline
p$_i$   &   $3.94$                   & $\pm 0.30$\\
              &   $1.18$                   & $\pm 0.05$  \\
\hline
B    (ppm$^2 \mu$Hz$^{-1}$)      &  0.91                    & $ \pm $0.011\\
\hline
\end{tabular}
\label{tab:campagne}
\end{table}

\section{p-mode oscillations parameters}
\label{seismo}
\subsection{Estimate of the large separation}
The Fourier spectrum of HD 46375 does not exhibit any seismic signature visible by eye (Fig. \ref{log_log}). To determine whether the star displays solar-like oscillations, we applied the method of \citet{Mosser_Appourchaux_2009}, based on the envelope autocorrelation function (EACF). This method searches for an oscillating signal in the autocorrelation of the time series, by analyzing the windowed Fourier spectrum proposed by \cite{Roxburgh_Vorontsov_2006}. It has shown to be efficient in low SNR cases, such as the CoRoT target HD 175726 \citep{Mosser_2009_175726}. The reliability of the result is given by an H0 test: when the EACF infers a signal above the threshold level, the null hypothesis can be rejected.

The method allows us, in a blind analysis without any a priori information, to derive the mean large separation of a solar-like spectrum. We were then able to derive an unambiguous detection around 155\,$\mu$Hz. The autocorrelation factor, around 11.5, is much above the threshold level of 8 calculated for a blind detection.

\subsection{Variation of the large separation with frequency}

\begin{figure}[t]
\includegraphics[width=9.2cm]{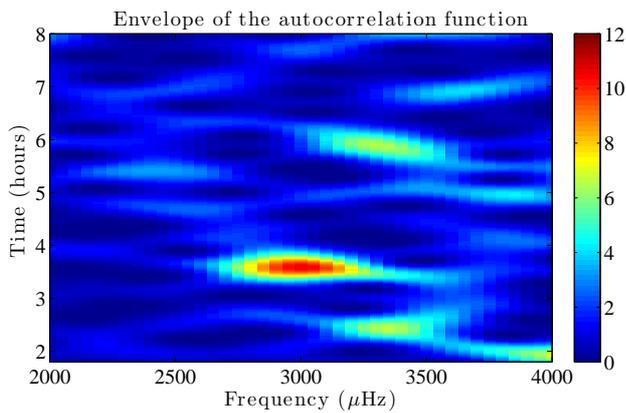}
\caption[]{Envelope of the autocorrelation function as function of the frequency and the time. The maximum value region, in the frequency range $[2600,3400]\ \mu$Hz, corresponds to a correlation time of about 3.6 h, that is to say a large separation of 153 $\mu$Hz. The EACF value is larger than 8 in the frequency range $[2750, 3300]\ \mu$Hz.}
\label{fig:EACF}
\end{figure}

\begin{figure}[t]
\includegraphics[width=8.3cm]{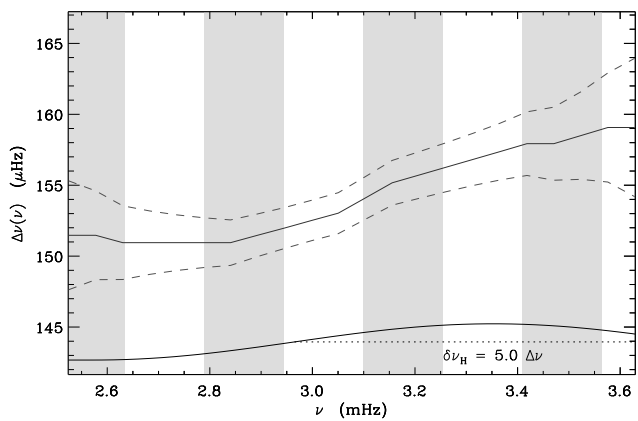}
\caption[]{Variation in the large separation with frequency, derived from the autocorrelation of the time series. The power spectrum was first windowed with a 0.75 mHz-half width cosine filter, as indicated in the lower-right corner. The sampling of the curve corresponds crudely to the mean large separation. The size of the dots is representative of the amplitude of the correlated signal. One-$\sigma$ error bars (dashed lines) are derived from \citet{Mosser_Appourchaux_2009}.}
\label{fig:autocorr}
\end{figure}

Measuring eigenfrequencies in the Fourier spectrum of  HD~46375 remains impossible, even by folding and collapsing the power spectrum in a collapsogram, to display a comb-like signal \citep{Korzennik_2008}. However, the autocorrelation method makes it possible to investigate the variation $\deltanunu$ in the large separation with frequency. The performance of the method depends on the autocorrelation factor. In the case of HD 46375, we can obtain 2 independent measures of the large separation, as indicated by Fig. \ref{fig:EACF}, from which we derive the large separation increase with frequency. The analysis of the spectrum  also provides error bars, as indicated in Fig. \ref{fig:autocorr}.


\subsection{p-mode amplitude}

The maximum bolometric amplitude of radial modes was derived according to \cite{Michel_2008}. The amplitude appears to be quite low, about $1.25 \pm 0.25$\,ppm. The large uncertainty originates first from the difficulty to estimate the background, and second from the very low height-to-background ratio $\HBR$, of about 2.5\,\%. 
This ratio measures the oscillation power relative to the power of the background, both intergrated over a 2-$\Delta\nu$ wide frequency range centered around the maximum oscillation frequency. The actual value of 2.5 \%
is very low, indeed much lower than the minimum value for which solar-like oscillations were derived in previous CoRoT observations (0.11 for HD 175726 according to \citealt{Mosser_2009_175726}). This shows the difficulty in measuring the large separation of HD~46375.

We can compare the observed amplitudes to the scaling law $(L/M)^{0.7}$ proposed by \citet{Samadi_2007}
for the Doppler signal. When translated into a photometric amplitude thanks to the adiabatic relation \citep{Kjeldsen_Bedding_1995}, this gives a scaling low $(L/M)^{0.7}/\sqrt{T\ind{eff}}$ according to which the bolometric amplitude of radial modes are estimated at the level of $A\ind{bol}(\ell=0) = 1.7$ ppm. As in most of the main-sequence stars observed by CoRoT, the observed amplitudes are significantly below the predictions.
Mode lifetimes cannot be measured. By comparing the frequency amplitudes with the autocorrelation signal, one can infer that mode lifetimes are longer at low frequency, as observed in other stars.

\section{Modelling}

We now construct a detailed asteroseismic model of HD~46375. The aim is to identify which stellar models are consistent with both the classical fundamental parameters ($T_{\rm eff}, \log g$, and metallicity) from Sect.~\ref{HR_spectro} and the seismic constraints from Sect.~\ref{seismo}.

\subsection{Seismic constraints}


\begin{table}
  \centering
  \caption{Classical and seismic parameters of HD~46375 derived from the observations
and the model that minimizes the $\chi^2$ function defined in Sect.
5.3.
  \label{tab_param}}
\begin{tabular}{c  c  c}
\hline
Parameters & Observations & Model \\
\hline \hline
\multicolumn{3}{l}{\textit{Observational constraints}} \\
\hline
$T\ind{eff} (K) $ & $5300\pm60$ & 5326 \\
$\log(L/L_{\odot})$ & $-0.17\pm0.04$ & $-0.22$ \\
$\log g$ & $4.66\pm0.09$ & 4.51 \\
($Z/X$)$\ind{s}$ & $0.060\pm0.008$ & 0.0626 \\
\hline
$\langle\Delta\nu\rangle$ & $153.0\pm0.7$ & 153.4 \\
\hline \hline
\multicolumn{3}{l}{\textit{Free parameters}} \\
\hline
Mass ($M_{\odot}$) & - & $0.97\pm0.05$ \\
Age (Gyr) & - & $ 2.6  \pm    0.8$ \\
$Y_0$ & - & $ 0.31 \pm  0.05$ \\
\hline
$\chi^2$ & - & 2.51\\
\hline
\end{tabular}
\end{table}

The autocorrelation method enabled us to obtain an estimate of the variation in the large separation with frequency. However, the frequencies at which we estimated $\Delta\nu$ depend on the width of the filter that we used, and the large separations cannot be directly compared to those of a model. We consequently used a mean value of the large separation $\langle\Delta\nu\rangle$ over the frequency domain where the EACF produces a signal above the threshold level. We also performed a linear regression upon $\Delta\nu$, taking into account the corresponding uncertainties given in Fig. \ref{fig:autocorr}. To perform a rapid but robust modeling, we reduced this variation to a linear trend, as in \citet{Mosser_2008_203608}. We therefore consider  (Table \ref{tab_param})
\begin{equation}
\Delta\nu=\Delta\nu_0+\gamma(\nu-\nu_0)
\end{equation}
The frequency $\nu_0=3$ mHz corresponds to the frequency of the maximum amplitude.

\subsection{Properties of the models}

Models were computed with the evolution code CESAM2k
\citep{Morel_1997}, and the mode frequencies were derived
from the models using the Liege OScillation Code (LOSC, see
\citealt{Scuflaire_2008}). We used the OPAL 2001 equation of state and opacity tables as described in \cite{Lebreton_2008}. Nuclear reactions rates were computed using the NACRE compilation \citep{Angulo_1999}. The models were calculated with the element abundances of  \cite{Grevesse_1993}.  We used Eddington's grey temperature-optical depth law to describe the atmosphere. The connection with the  envelope was achieved at an optical depth of 10. 
Convection was described using the Canuto, Goldman, and Mazzitelli (CGM, \citealt{Canuto_1996}) formalism, which involves a free parameter assumed to be some fraction $\alpha\ind{CGM}$ of the local pressure scale height $H_p$. In our study, this parameter was fixed to the solar value ($\alpha\ind{CGM} = 0.64$). 

Overshooting is described as an extension of the convective core over a distance $d\ind{ov}=\alpha\ind{ov} \min(r\ind{core},H_p)$, where $r\ind{core}$ is the radius of the convective core. Microscopic diffusion is not taken into account.

In contrast to heavier elements, it is difficult to determine spectroscopically the helium content $Y$ of the stellar surface. Hence, $Y$ is assumed to be a free parameter of the models. However, several studies have shown that a link exists between the helium enrichment of the galaxy and the enrichment of heavier metals. This leads to the determination of a helium-to-metal enrichment ratio, written $\Delta Y/\Delta Z$ such that
\begin{equation}
Y=Y\ind{P}+\frac{\Delta Y}{\Delta Z}\ Z.
\label{eq_yp}
\end{equation}
For K-type stars around and above solar metallicity, \cite{Casagrande_2007} found $\Delta Y/\Delta Z\sim 2.1 \pm 0.9$. 
We can thus derive an estimate of $Y$. Inserting Eq. \ref{eq_yp} into $X+Y+Z=1$, one indeed obtains
\begin{equation}
Y=1-\frac{1-Y\ind{P}}{1+(Z/X)\left( 1+\Delta Y/\Delta Z \right)}\ \left[ 1+ (Z/X)  \right]
\label{eq_y}
\end{equation}
Taking $Y\ind{P}=0.248\pm0.003$ (see \citealt{Peimbert_2007}), and using our spectroscopic determination of the metallicity $(Z/X)=0.063\pm0.006$, we find that $Y=0.33\pm0.05$. 

\subsection{Optimization}
\label{modele}
We now search for models that reproduce the observational constraints (the
effective temperature $T\ind{eff}$, the luminosity $L$, the surface
gravity $\log g$, the metallicity $(Z/x)\ind{s}$, and the mean large
separation $\langle\Delta\nu\rangle$) using as free parameters the stellar
mass, age, and initial content in helium $Y_0$. The values of the
observational constraints are summarized in Table 3.

To estimate the compatibility of the models with the observations, we use
\begin{equation}
\chi^2=\sum_{i=1}^{N} \left( \frac{\lambda_i\ex{obs}-\lambda_i\ex{mod}}{\sigma_i\ex{obs}} \right)^2
\end{equation}
where $\lambda_i\ex{obs}$ is the $i\ex{th}$ observational constraint (among $N$), $\sigma_i\ex{obs}$ its 1-$\sigma$ error bar, and $\lambda_i\ex{mod}$ the value of the corresponding parameter in the considered model.

To search for an optimal model, we use the Levenberg-Marquardt algorithm, as described by \cite{Miglio_Montalban_2005}. The interest of this method is that it enables us to minimize a function with a small number of iterations, even with initial parameters rather far away from the final result. However, when using this optimization method, we must pay attention to the risk of converging to a local, not global, minimum. Hence, we combined this approach with the computation of a grid of models.

The parameters of the optimal model we found are given in Table 3. All the
classical parameters are fitted within 1-$\sigma$ error bars, except the
stellar luminosity for which the obtained value differs from the observed
one by 1.4 $\sigma$. The mean value of the large separation is reproduced well by our best-fit model. We however note that if we perform a linear
fit to the variations in the large separation for this model, we obtain a
slope of $\gamma\ind{mod}=-1.1\;10^{-4}$, quite different from the one we
derive with the autocorrelation of the time series ($0.010 \pm 0.003$). To investigate this disagreement, individual frequencies would be
required.
This would require more complex modelling based on observations in higher SNR, as should be provided with the reobservation of this exoplanet-hosting star.

\section{Discussion and conclusion}

This work is particularly interesting because of its overlap between exoplanetary science and asteroseismology, as also performed by the {\it Kepler} mission \citep{Christensen-Dalsgaard_2010} and the proposed {\it Plato} mission. Even for very low signal-to-noise ratio data, as presented here, we have demonstrated that we can constrain the stellar fundamental parameters and hence also the planetary parameters. Moreover, the ground-based support data provided additional constraints on the magnetic field as do high resolution spectra, which have been used to refine the atmospheric parameters.

From the Zeeman Doppler imaging observations, performed with the NARVAL spectropolarimeter at the Pic du Midi Observatory, we reconstructed the stellar magnetic structure. We found that the magnetic field of HD~46375 is significantly different from other planet-bearing stars, such as $\tau$ Boo and HD~189733: for HD~46375, the magnetic field is mainly poloidal, while it is predominantly toroidal for the two other stars. This difference is related to the stellar rotation. Using the rotation-activity relationship, the stellar rotation period is $42^{+9}_{-7}$ day. Moreover, since the magnetic field measurements do not display any periodicity, we have confirmed that the period is longer than the run-duration (35 days). With this rotation period, HD~46375 is among stars featuring very simple, mostly poloidal, and axisymmetric large-scale magnetic topologies.

The comparison of the fundamental parameters before and after this work illustrates the potential of asteroseismology. The stellar mass has increased by about 7 \% and the stellar radius by the same proportion, which corresponds to a much denser star with a correspondingly higher large separation. Despite the low signal, the running filtered autocorrelation method \citep{Mosser_Appourchaux_2009} allows us to measure the large separation of the p-mode frequencies. The mean value of the large separation is $153.0 \pm 0.7$ $\mu$Hz, which is the highest value of the large separation ever observed in photometry. Only three targets with higher $\Delta\nu$ have been observed in spectroscopy ($\tau$~Ceti, $\alpha$~Cen~B, and 70~Oph~A). The Fourier analysis of the CoRoT light curve exhibits a tiny excess of power, with amplitude of about 1.25 ppm, i.e., a height-to-background ratio of about 2.5\%. 

We combine seismic parameters with the atmospheric parameters obtained from the high resolution NARVAL spectra. This allows us to refine the fundamental parameters of the star i.e., its mass, effective temperature, radius, and age. The best-fit model has a mass of $0.97\pm0.04\ M\ind{\sun}$ with an age of $2.6\pm0.8$ Gyr. Despite the relatively large error bar in the age, all the other inputs, such as the high metallicity, the lower temperature, and luminosity with respect to the Sun reinforce the hypothesis of a young metal-rich star. Thus, it is not exactly a solar twin, but a young Sun with a composition enhanced in metallic elements. As expected from the asteroseismic diagram proposed by \citet{Christensen-Dalsgaard_1988}, this younger analog of the Sun has a higher value of the large separation. However, since we have not been able to identify the individual eigenfrequencies, we cannot constrain the age more tightly.

The revision of the stellar mass permits us to re-estimate the projected mass of the known exoplanet to be $M\ind{p} \sin i = 0.234 \pm 0.008\ M\ind{jup}$, instead of $0.226\pm0.019\ M\ind{jup}$ in \citet{Butler_2006}. Although we claim an improvement in the accuracy by a factor two, we cannot constrain the true planetary mass. If we assume the planet orbital plane is in the stellar equatorial plane, the planetary mass is $M\ind{p} = 0.29^{+0.14}_{-0.06}\ M\ind{jup}$. However, we do not know whether the orbital plane is in the stellar equatorial plane as in the Solar System: several exo-planetary systems do show different configurations (e.g. \citealt{Winn_2009}, \citealt{Hebrard_2008}).

The CoRoT short run on HD~46375 was part of an additional program: it has demonstrated that we are able to carry out a complete analysis of a stellar system when combining photometry and ground-based spectrometric support. However, a 34-day run appears too short to provide high-quality data for such a relatively faint star. From a 5-month long run with CoRoT, it would be possible to derive individual eigenfrequencies and to monitor in more detail the rotation period. Moreover, a longer run should be able to confirm the detection of a planetary contribution to the light curve, to be reported in our forthcoming Paper~II.


\begin{acknowledgement}
P. Gaulme acknowledges financial support from the Centre National d'Etudes Spatiales (CNES) for a post-doctoral fellowship. Part of this work is based on observations obtained with NARVAL at the T\' elescope Bernard Lyot (TBL), operated by INSU/CNRS. Finally, we warmly thank John Leibacher for his patience in carefully reading the paper.
\end{acknowledgement}

\bibliographystyle{aa} 
\bibliography{bibi} 

\end{document}